\documentclass[9pt,conference]{IEEEtran}
\IEEEoverridecommandlockouts
\usepackage{cite}
\usepackage{amsmath,amssymb,amsfonts}
\usepackage{algorithmic}
\usepackage{graphicx}
\usepackage{textcomp}
\usepackage{xcolor}
\usepackage[hyphens]{url}  
\usepackage{hyperref}
\usepackage{tabularx}   
\usepackage{booktabs}
\usepackage{graphicx} 
\usepackage{multirow}

\usepackage{colortbl}   
\usepackage{xcolor}     

\definecolor{bg_gray}{gray}{0.92}

\def\BibTeX{{\rm B\kern-.05em{\sc i\kern-.025em b}\kern-.08em
    T\kern-.1667em\lower.7ex\hbox{E}\kern-.125emX}}
\begin{document}

\title{Semantic-Aware Interruption Detection in Spoken Dialogue Systems: Benchmark, Metric, and Model}


\author{Kangxiang Xia\textsuperscript{1}, Bingshen Mu\textsuperscript{1}, Xian Shi\textsuperscript{1}, Jin Xu\textsuperscript{1*}\thanks{* Corresponding Author. This work was done when Kangxiang Xia was an intern in Qwen Team, Alibaba, Beijin, China.}, Lei Xie\textsuperscript{2}\\
        \textsuperscript{\rm 1}Qwen Team, Alibaba\\
    \textsuperscript{\rm 2}Independent Researcher\\
    }

\maketitle

\begin{abstract}
Achieving natural full-duplex interaction in spoken dialogue systems (SDS) remains a challenge due to the difficulty of accurately detecting user interruptions. Current solutions are polarized between "trigger-happy" VAD-based methods that misinterpret backchannels and robust end-to-end models that exhibit unacceptable response delays. Moreover, the absence of real-world benchmarks and holistic metrics hinders progress in the field. This paper presents a comprehensive framework to overcome these limitations. We first introduce SID-Bench, the first benchmark for semantic-aware interruption detection built entirely from real-world human dialogues. To provide a rigorous assessment of the responsiveness-robustness trade-off, we propose the Average Penalty Time (APT) metric, which assigns a temporal cost to both false alarms and late responses. Building on this framework, we design an LLM-based detection model optimized through a novel training paradigm to capture subtle semantic cues of intent. Experimental results show that our model significantly outperforms mainstream baselines, achieving a nearly threefold reduction in APT. By successfully resolving the long-standing tension between speed and stability, our work establishes a new state-of-the-art for intelligent interruption handling in SDS. To facilitate future research, SID-Bench and the associated code are available at: https://github.com/xkx-hub/SID-bench.
\end{abstract}


\section{Introduction}
\label{sec:intro}

The pursuit of human-like, fluent spoken interaction is a long-standing goal in the field of Spoken Dialogue Systems (SDS)~\cite{GoogleDuplex,Qwen2.5-Omni, AudioGPT, SpeechGPT, OSUM-EChat, SALMONN_omni, NTPP}. A critical milestone toward this goal is full-duplex technology, which allows systems to ``listen" and ``speak" simultaneously~\cite{moshi, MinMo, LM_can_listen_while_speaking, beyound_turn_based}. However, achieving bidirectional transmission at a technical level is not the final destination. The faithful measure of interactional naturalness lies in a system's capacity to handle dynamic, non-sequential conversational behaviors~\cite{Turn-taking-cues, ssr-in-con, Optimizing-turn-taking}. Among these, intelligent handling of user interruptions is of paramount importance, as it elevates an interaction from a rigid, turn-taking model to a collaborative, efficient real-world conversation.

Despite the clear importance of interruption detection, current solutions often fall short. Mainstream full-duplex systems rely on Voice Activity Detection (VAD) to identify interruptions~\cite{Freeze-Omni, vita}. While fast, these systems are often ``trigger-happy", they frequently mistake simple user backchannels (like ``uh-huh" or ``right") or brief hesitations as signals to stop, causing the system to cut itself off inappropriately. On the other end of the spectrum, some end-to-end models~\cite{moshi} prioritize robustness but suffer from high latency, failing to stop promptly even when the user clearly intends to take the floor. At their core, existing systems lack a deep understanding of the user’s underlying intent, making interactions feel mechanical and clumsy.

A more severe challenge than the technical implementation itself is the absence of adequate evaluation methods. Methodologies for assessing interruption capabilities have lagged behind technological advancements. 
Human evaluation, while nuanced, is too costly, subjective, and difficult to replicate for rapid system development~\cite{Full-Duplex-Bench-v2}.
To address this, fully automated benchmarks such as full-duplex-bench series~\cite{Full-Duplex-Bench, Full-duplex-benchv1.5, Full-Duplex-Bench-v2} and FD-Bench~\cite{FD-Bench} have recently emerged, making significant strides in the systematic evaluation of overlapping speech. 
However, these advanced works still exhibit several key limitations. Fundamentally, both rely on synthetic data, text generated by LLMs and spoken by TTS.
While scalable, this approach fails to capture the intricate blend of timing, prosody, and emotion found in spontaneous human speech, leaving a gap in ``real-world" realism\cite{turn_taking_review, pause_gaps_overlaps}.
Furthermore, existing automated metric systems are either too complex to be easily interpretable or lack a unified composite score to balance different performance dimensions, such as responsiveness versus robustness. 
Moreover, these works primarily focus on ``diagnosing the problem" rather than providing a validated, constructive solution to enhance the model's interruption capabilities. Therefore, a framework based on real-world data, featuring a concise and compelling evaluation system, and capable of guiding model optimization, remains a significant gap in the field.

To bridge these dual gaps in evaluation and model capabilities, this paper presents the following contributions. First, we introduce the Semantic Interruption Detection Benchmark (SID-Bench), a novel benchmark constructed from real-world, recorded human conversations, designed explicitly for evaluating interruption handling. Unlike methods reliant on synthetic data, SID-Bench captures the diverse and spontaneous interruption patterns found in authentic interactions. Second, to enable a precise measurement of interruption intelligence, we define a new set of evaluation metrics focusing on accuracy and latency. 
To address their intrinsic trade-off, we propose a composite score that provides a unified performance indicator by encapsulating the penalties associated with both false alarms and delayed responses.
Third, to establish a strong baseline on this benchmark, we propose a novel LLM-based model for interruption detection. By employing a 'large-scale pre-training then few-shot fine-tuning' paradigm, this approach leverages pre-existing knowledge to acquire specialized interruption detection skills without requiring extensive task-specific annotated data.
Finally, we conduct a comprehensive evaluation on SID-Bench, benchmarking our proposed model against several mainstream full-duplex systems. Experimental results demonstrate that our model significantly outperforms existing systems, proving not only the effectiveness of our approach but also validating the efficacy and necessity of SID-Bench in differentiating the interruption capabilities of various models.

\section{SID-bench}
To address the critical gap in ecologically valid evaluation, we introduce the SID-Bench, a novel benchmark meticulously designed to assess the interruption-handling capabilities of SDS. In contrast to prior works that rely on synthetic data, SID-Bench is built upon authentic, recorded human-human conversations, ensuring that the evaluation scenarios accurately reflect the complex and spontaneous nature of real-world interactions. Its core contribution lies in providing an evaluation framework that not only measures a system's responsiveness but also, crucially, its ability to differentiate between genuine interruptions and non-interruptive backchannels. This section outlines the data collection, annotation methodology, and statistical properties of the resulting benchmark.

\begin{figure}[t]
\centering
\includegraphics[width=1.0\linewidth]{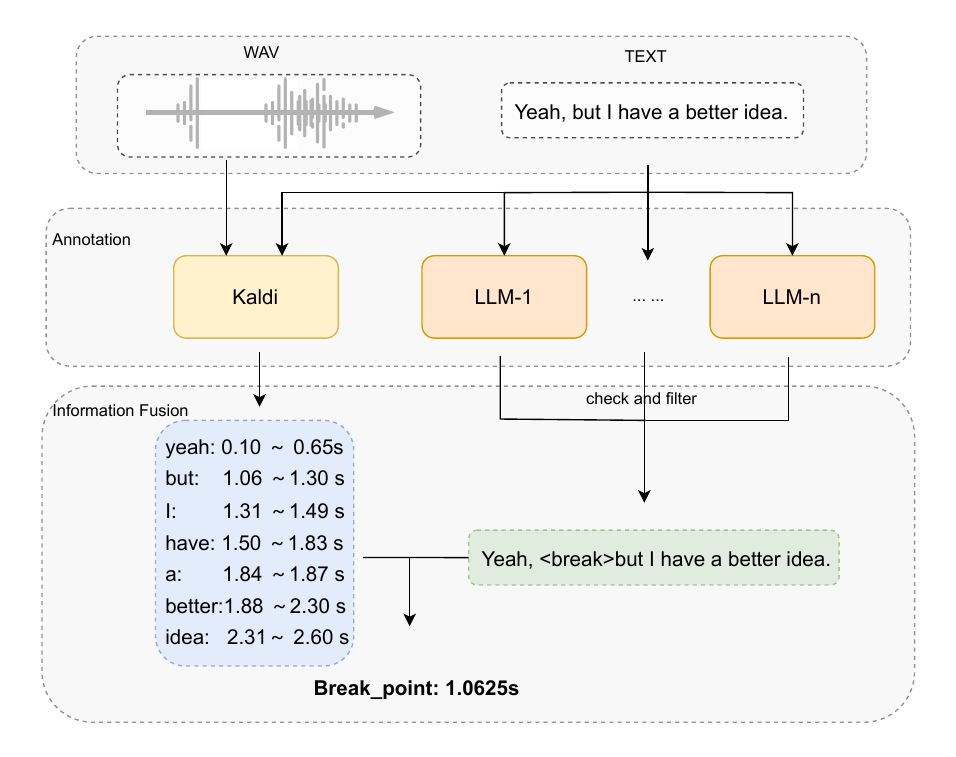}
\vspace{-6pt}
\caption{The semi-automated annotation pipeline for SID-Bench. The process begins with raw audio and its transcription. In the Annotation stage, the audio is processed by the Kaldi toolkit for forced alignment to obtain word-level timestamps, while the text is analyzed by a series of LLMs to semantically identify the interruption point, marked with a \textless break\textgreater~tag. In the final Information Fusion stage, the semantic \textless break\textgreater~tag is aligned with the precise start time of the corresponding word from Kaldi, establishing a semantically meaningful and temporally accurate ground-truth Break\_point.}
\label{fig:data_piple}
\end{figure}

\subsection{Data Collection and Sources}
The foundation of SID-Bench is the use of real-world conversational data, as we believe that the subtle acoustic and semantic cues inherent in human interruptions can only be captured through authentic recordings. The benchmark is bilingual, comprising both Chinese and English conversations. The Chinese data was professionally recorded by a data vendor, featuring natural dialogues. The English portion combines similarly recorded vendor data with a curated subset from the widely-recognized Switchboard~\cite{SWITCHBOARD} corpus, a standard resource known for its spontaneous conversational style.

The final benchmark comprises a total of 3,700 evaluation instances, extracted from approximately 10 hours of curated audio.  The included conversational turns are not pre-selected based on whether they contain an interruption; instead, they represent a natural distribution of overlapping speech events found in human-human dialogue. SID-Bench considers genuine interruption, simple backchannels, and ambiguous cases, creating a realistic and challenging testbed for a model's judgment.

\subsection{Definition and Annotation of Interruption Events}
A primary limitation of existing systems is their inability to distinguish between overlapping speech and true interruption intent. To facilitate a more nuanced evaluation, we establish a clear, functional definition for interruption events. We define a true Interruption as an event where a speaker begins their turn with a new communicative intent that semantically warrants the other speaker to yield their turn. In contrast, a Backchannel is an utterance used to acknowledge, agree, or show continued attention while not altering the conversational goal, exemplified by expressions such as ``uh-huh," ``right," and ``I see."

To accurately and consistently label the precise moment of interruption, we developed a semi-automated annotation pipeline, as illustrated in Figure~\ref{fig:data_piple}. This pipeline uniquely combines the semantic analysis of LLMs with high-precision temporal information from forced alignment. First, we prompt powerful LLMs, Qwen-max, Qwen-plus~\cite{Qwen2.5}, to analyze the raw text of a speaker's turn and insert a \textless break\textgreater~tag at the exact word where the utterance transitions from non-substantive backchanneling to expressing a new, substantive intent. To ground this semantic breakpoint in the time domain, we utilize the Kaldi toolkit to perform forced alignment on the raw audio, thereby generating precise timestamps for each word and phoneme. By integrating these two stages, we define the ground-truth interruption time as the start time of the first phoneme of the first substantive word identified by the LLM. This hybrid approach ensures our labels are not only temporally precise but also semantically meaningful, directly reflecting the speaker's underlying intent.

\begin{table}[t]
\centering
\caption{The composition and statistics of SID-bench dataset.}
\label{tab:dataset_stats}
\begin{tabular}{llc}
\toprule
\textbf{Language} & \textbf{Category} & \textbf{Count} \\
\midrule
\multirow{3}{*}{Chinese} & Interruption at the beginning & 500 \\
                              & Interruption in the middle & 600 \\
                              & Uninterrupted & 500 \\
\midrule
\multirow{3}{*}{English} & Interruption at the beginning & 500 \\
                              & Interruption in the middle & 600 \\
                              & Uninterrupted & 500 \\
\midrule
\multirow{2}{*}{Noise/Silence} & Environmental noise & 200 \\
& Silence with subtle sounds & 300 \\
\bottomrule
\end{tabular}
\end{table}

\subsection{Statistics of SID-Bench}
Table~\ref{tab:dataset_stats} outlines the composition of SID-Bench, a benchmark designed for the comprehensive evaluation of interruption handling, including three main data categories.

The collection of 3,200 conversational instances is evenly split between Chinese and English. These instances are further classified based on the user's overlapping speech patterns. Among them, 500 instances are designated as \textit{Uninterrupted}, containing only non-substantive backchannels or filler words without any intent to take the conversational turn. The other 1,100 instances represent genuine interruptions, which are further partitioned into 500 \textit{Interruption at the Beginning} cases, where intent is immediately apparent, and 600 \textit{Interruption in the Middle} cases, where an initial backchannel evolves into a substantive interjection. The detailed categorization enables a rigorous assessment of a model's intent differentiation performance. Furthermore, the Noise and Silence set of 500 instances is included to test the system's robustness against non-semantic acoustic phenomena. This set is composed of 300 Silence clips and 200 Noise clips. The Silence clips are synthesized by concatenating near-silent segments from original recordings to a duration of 10 seconds; these segments might include low-level audio artifacts like breathing or microphone hiss. The Noise clips contain prevalent environmental sounds, such as rain and keyboard typing. This set is vital for verifying that the model responds specifically to semantic speech cues rather than reacting to any arbitrary acoustic energy.

\section{Evaluation Metrics}
\begin{figure}[ht]
\centering
\includegraphics[width=1.0\linewidth]{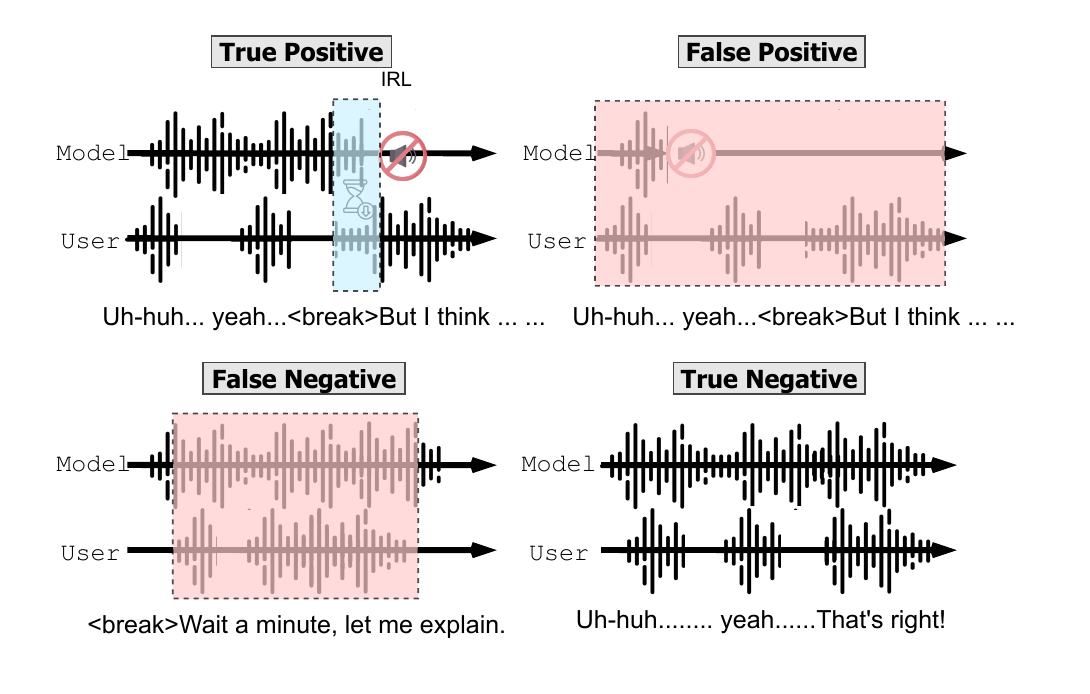}
\vspace{-6pt}
\caption{
Illustration of the four evaluation scenarios and their associated time penalties. The user's utterance contains a ground-truth interruption intent marked by  \textless break\textgreater~.
\textbf{(a) True Positive:} The system correctly stops after the  \textless break\textgreater~ point. The penalty is IRL, shown in blue.
\textbf{(b) False Positive:} The system incorrectly stops in response to a backchannel before the \textless break\textgreater~. This is a catastrophic failure, and the penalty, shown in red, applies to the entire turn's duration.
\textbf{(c) False Negative:} The system fails to stop, forcing the user to listen to superfluous speech. The penalty, shown in red, is the duration of this unwanted audio.
\textbf{(d) True Negative:} The system correctly ignores a backchannel and continues speaking when no \textless break\textgreater~ is present, thus incurring zero penalty.
}
\label{fig:metrics}
\end{figure}

A quantitative assessment of what we term ``interruption intelligence" necessitates an evaluation framework that transcends traditional metrics. The core challenge in designing such a system is resolving the inherent trade-off between two competing demands. The first is accuracy, the system's ability to prevent spurious interruptions. The second is timeliness, its capacity to respond promptly to valid interruptions. To formally capture this balance, we introduce a suite of metrics. These include two primary indicators adopted from~\cite{FD-Bench}, the False Interruption Rate (FIR) and the Interruption Response Latency (IRL). To provide a holistic measure, we also formulate a single composite score, the Average Penalty Time (APT), which synthesizes the penalties associated with both false alarms and delayed responses. Our entire metric set is systematically derived from the four distinct outcomes of user-system interaction, detailed in Figure~\ref{fig:metrics}.

\begin{figure*}[htp]
\centering
\includegraphics[width=0.88\linewidth]{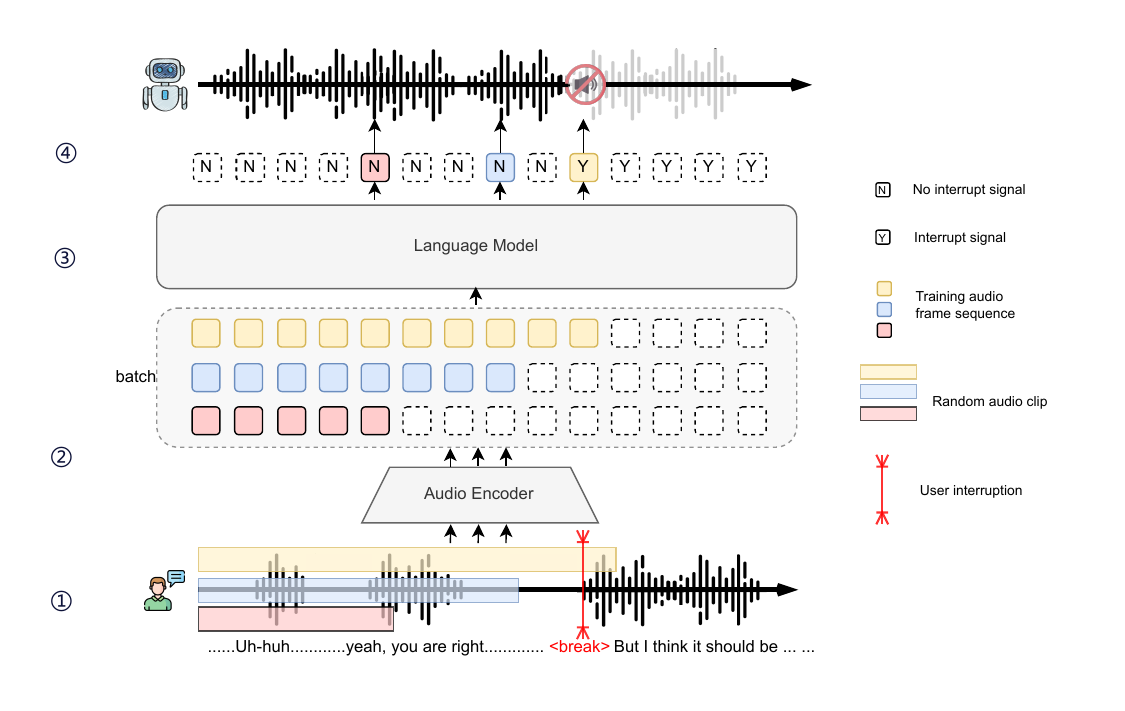}
\caption{The overall architecture of the proposed SID-model for real-time interruption detection.
(1)  During training, we use audio with a pre-labeled user interruption point. This audio is randomly cropped into clips of varying lengths, represented by different colors. Each clip is assigned a ground-truth label: `Y', interrupt, if its endpoint is after the user interruption point, and `N' otherwise.
(2) The audio clips are processed by an Audio Encoder (AuT) to extract features. 
(3) The resulting audio frame sequences are fed into a LLM (Qwen3-0.6b).
(4) The model sequentially predicts whether to issue an interrupt signal or not, enabling the system to stop its speech in a timely manner. The user interruption points are annotated following the procedure of the SID-bench.
}
\label{fig:methodology}
\end{figure*}

\subsection{Primary Metrics: FIR and IRL}
\textbf{FIR} measures a system's robustness. It is the proportion of test cases in which a system incorrectly stops its utterance in response to a non-interruptive vocalization, such as a backchannel. A lower FIR indicates a less ``trigger-happy" and more natural-sounding system. FIR can be calculated as:
\begingroup
\begin{align}
\text{FIR} = \frac{\text{Number of False Interruptions}}{\text{Total Number of Test Cases}}.
\end{align}
\endgroup

\textbf{IRL} measures a system's responsiveness. For correctly handled interruptions (True Positives), it is the time elapsed from the ground-truth interruption intent $T_{break}$ to the system's stop time $T_{model\_stop}$. A lower IRL signifies a more agile system. IRL can be calculated as:
\begingroup
\begin{align}
\text{IRL} = T_{\text{model\_stop}} - T_{\text{break}}.
\end{align}
\endgroup

\subsection{Composite Score: APT}

To provide a single, holistic measure, we introduce APT. This score combines accuracy and timeliness by assigning a ``time cost" equivalent to every system action for the user, as visualized in Figure~\ref{fig:metrics}. The penalty for each outcome is defined as follows:
\begin{itemize}
\item \textbf{True Positive:} The penalty is the unavoidable latency, \textbf{IRL}.
\item \textbf{False Positive:} A catastrophic failure that derails the conversation. The penalty is therefore the \textbf{total planned duration of the turn}, representing a complete loss of that interactional opportunity.
\item \textbf{False Negative:} Force the user to listen to unwanted speech. The penalty is the \textbf{duration of this superfluous audio} after the user has expressed their intent.
\item \textbf{True Negative:} An ideal behavior incurs \textbf{zero penalty}.
\end{itemize}

The final APT score is the average penalty across all $N$ test cases, providing a powerful and intuitive measure of overall performance:
\begin{equation}
    \text{APT} = \frac{1}{N} \sum_{i=1}^{N} \text{PenaltyTime}_i,
    \label{eq:apt}
\end{equation}
a lower APT signifies a superior system that better balances responsiveness and robustness.

\section{Methodology}
To facilitate precise, real-time identification of user interruption intent, we propose a framework built upon a large-scale multimodal architecture. The core of our approach lies in reframing interruption detection as a sequential classification problem. Through a specialized two-stage training paradigm and a robust inference strategy, the model learns to generate timely decisions grounded in a semantic understanding of the user’s utterance.

\subsection{Model Architecture}

Our primary objective is to move beyond VAD toward a deep semantic comprehension of conversational intent. As illustrated in Figure~\ref{fig:methodology}, our model comprises two primary components: an audio encoder and a LLM decoder.
The architecture is inspired by state-of-the-art multimodal designs such as Qwen3-Omni~\cite{Qwen3-Omni}, aiming for a seamless fusion of acoustic features and linguistic semantics. We employ an Audio Transformer (AuT) as the audio encoder to extract high-dimensional acoustic embeddings from raw signals. These embeddings are then fed into Qwen-0.6b~\cite{Qwen3}, a lightweight yet powerful LLM that serves as the central reasoning engine. By leveraging its advanced contextual modeling capabilities, the LLM processes the sequence of feature frames and generates a binary classification output, determining whether the accumulated audio context constitutes a genuine interruption intent.

\subsection{Training Paradigm}
To efficiently teach the model ``when to interrupt", we designed a two-stage ``pre-training for alignment, fine-tuning for detection" paradigm, augmented with several training strategies tailored to the interruption task.

The initial phase aims to bridge the representation gap between acoustic and linguistic modalities. We pre-train the entire pipeline on large-scale ASR datasets. By optimizing for the ASR task, the model is forced to map acoustic signals onto their corresponding textual units, ensuring that the audio encoder's output can be effectively interpreted by the LLM’s language space. This stage establishes a robust multimodal foundation for subsequent intent detection phase.

In the second phase, the model is fine-tuned specifically for the interruption detection task. We utilize a bilingual dialogue corpus processed similarly to SID-Bench, providing a training set with precise interruption timestamps $T_{break}$.
As depicted in Figure~\ref{fig:methodology}, training samples are generated using a randomized temporal cropping method. For any given utterance containing a ground-truth interruption $T_{break}$ , we extract segments of varying lengths. Each segment is assigned a binary label based on its endpoint relative to $T_{break}$ : a segment is labeled as a positive instance if its endpoint exceeds $T_{break}$, and a negative instance otherwise. This label serves as the supervision signal for the LLM’s final hidden state, training the model to evaluate whether the growing audio context provides sufficient evidence of an interruption intent.

To address the inherent ambiguity and temporal sensitivity of interruption behaviors, we incorporate two specialized boundary-aware training strategies: intent confirmation delay and boundary-aware sampling. 
First, in authentic interactions, confirming an intent often requires a brief context following the initial trigger word. To simulate this, we apply a one-word temporal shift  to the ground-truth $T_{break}$ during label generation. This forces the model to integrate more evidence before issuing a positive prediction, thereby enhancing robustness against premature and erroneous interruptions.
Second, the interval immediately surrounding $T_{break}$ represents the most critical decision-making window. To ensure the model effectively learns these subtle distinctions, we increase the sampling density of clips whose endpoints fall near this boundary. Specifically, for each source utterance, we guarantee that at least two sampled clips span opposite sides of $T_{break}$, forcing the model to discriminate between non-substantive backchannels and the onset of substantive intent.


\begin{table*}[ht]
\centering
\caption{Comparison of different models on Average IRL, False Interruption Rate, and Average Penalty Time across EN, ZH, and Noise conditions. Lower values are better (indicated by $\downarrow$).}
\label{tab:metrics_comparison}

\renewcommand{\arraystretch}{1.2} 
\setlength{\tabcolsep}{3pt} 
\footnotesize

\begin{tabular*}{\textwidth}{@{\extracolsep{\fill}} l ccc ccc cc ccc @{}}
\toprule
& \multicolumn{3}{c}{\textbf{EN}} & \multicolumn{3}{c}{\textbf{ZH}} & \multicolumn{2}{c}{\textbf{Noise Silence}} & \multicolumn{3}{c}{\textbf{Average}} \\
\cmidrule(lr){2-4} \cmidrule(lr){5-7} \cmidrule(lr){8-9} \cmidrule(lr){10-12}
\textbf{Model} & IRL$\downarrow$ & FIR$\downarrow$ & APT$\downarrow$ & IRL$\downarrow$ & FIR$\downarrow$ & APT$\downarrow$ & FIR$\downarrow$ & APT$\downarrow$ & IRL$\downarrow$ & FIR$\downarrow$ & APT$\downarrow$ \\
\midrule
FSMN-VAD~\cite{FunASR}& \textbf{0} & 0.906 & 3.328 & \textbf{0} & 0.915 & 3.563 & 0.392 & 4.791 & \textbf{0} & 0.840 & 3.627 \\
Free-Omni(Silero VAD)~\cite{Freeze-Omni} & \underline{0.141} & 0.669 & 2.429 & \underline{0.186} & 0.513 & \underline{1.977} & 0.154 & \underline{1.658} & \underline{0.173} & 0.532 & \underline{2.129} \\
FireRedChat (pVAD)~\cite{FireRedChat} & 0.474 & 0.734 & 2.997 & 1.170 & \underline{0.338} & 2.311 & \underline{0.098} & 2.458 & 1.045 & 0.476 & 2.627 \\
Moshi~\cite{moshi} & 2.517 &\textbf{ 0.010} & \underline{1.749} & - & - & - & 0.464 & 7.810 & 2.517 & \textbf{0.118} & 3.192 \\
SID-model (proposed) & 0.444 & \underline{0.140} & \textbf{0.921} & 0.338 & \textbf{0.138} & \textbf{0.656} & \textbf{0.026} & \textbf{0.215} & 0.389 & \underline{0.124} & \textbf{0.711} \\
\bottomrule
\end{tabular*}
\end{table*}

\subsection{Inference Process}

To emulate a real-world streaming dialogue scenario, our inference process operates on a continuous stream of user audio. The incoming audio is segmented into fixed-length chunks. At each step, these chunks are cumulatively concatenated and fed into the model (e.g., $\text{chunk}_{1}$, $\text{chunk}_{1+2}$, $\text{chunk}_{1+2+3}$, ...), allowing the model to make predictions based on all available historical context.
However, the acoustic and semantic signals around an interruption boundary are often ambiguous, which can cause the model's predictions to be unstable and oscillate between Continue and Interrupt. To mitigate the risk of false positives triggered by this momentary prediction jitter, we implement a decision smoothing policy. A definitive interruption event is registered, and a stop command is issued to the dialogue system, only after the model outputs Interrupt for $K$ consecutive steps. This hyperparameter $K$ allows for a direct trade-off between responsiveness and decision stability. In our experiments, we set $K = 3$.

\section{Experiments}
We applied our proprietary SID-Bench benchmark to evaluate the open-source FDSDS models, which serve as strong performance baselines. This process is designed to validate the efficacy of our proposed model and the diagnostic power of SID-Bench itself.

\subsection{Baselines}

 We compare our approach against four representative open-source systems, each reflecting a mainstream technical paradigm for interruption handling: (1) \textbf{FSMN-VAD}~\cite{FunASR} is an efficient VAD model that represents the most fundamental interruption strategy, which treats any detected user speech as an interruption trigger. (2) \textbf{Freeze-Omni (Silero-VAD)}~\cite{Freeze-Omni} is an open-source cascaded full-duplex system that combines a frozen LLM with chunked streaming audio input and turn-taking management. User input detection is handled by the widely used Silero-VAD, representing a typical deployment of VAD in SDS. (3) \textbf{FireRedChat (pVAD)}~\cite{FireRedChat} is another advanced, open-source, full-duplex system. Its key component is a proprietary VAD, specifically engineered for robustness against non-interruptive sounds, which enables a more optimized VAD-based approach. (4) \textbf{Moshi}~\cite{moshi} is an end-to-end, real-time speech-to-speech model. It employs an ``Inner Monologue'' mechanism and a multi-stream design to improve fluency and handle overlapping speech, exemplifying the capabilities of fully integrated end-to-end architectures.

\subsection{Evaluation} 

To validate our proposed model and establish the utility of our new benchmark, we conduct a rigorous comparative analysis.
All systems are evaluated on SID-Bench using three key metrics from Section 3: 1) FIR to quantify robustness against spurious triggers; 2) IRL to measure timeliness; and 3) APT, our proposed composite metric that provides a single, holistic score of a system's ability to balance the trade-off between accuracy and timeliness.

Given the significant architectural and input-output differences across these systems, we establish a unified evaluation protocol on SID-Bench focused solely on measuring the response time aligned with the user's interruption intent.
For the VAD-based baselines, we take the onset time detected by the corresponding VAD as the model's \emph{user barge-in time point}, and compute all metrics with respect to this timestamp.
For the end-to-end model Moshi, which does not expose an explicit barge-in signal, we design a dedicated procedure. While Moshi is generating a response to a preset prompt, we start playing a SID-Bench test audio file at a fixed time. We fix the random seed to ensure a deterministic initial response from Moshi. We then record the timestamp at which Moshi stops its ongoing output, and treat the onset of our test audio playback as the \emph{user barge-in time point} for evaluation. This protocol allows us to consistently assess Moshi's end-to-end responsiveness.

\subsection{Main Results and Analysis}
The comparative performance of our model against the baselines across all test conditions is presented in Table \ref{tab:metrics_comparison}. The results demonstrate that our proposed model achieves a dominant advantage in the composite APT metric, and is the only model to strike an ideal balance between low IRL and low FIR.

\textbf{Pure VAD Model (FSMN-VAD): Fast but Fundamentally Flawed.} The FSMN-VAD baseline exemplifies a ``trigger-happy" strategy, reacting instantaneously to any detected vocal energy. This results in a nominal IRL of 0 but at the cost of an extremely high FIR, which averages 0.840 and exceeds 0.90 in conversational contexts. Such an approach, while maximally responsive, lacks the discernment to distinguish genuine interruptions from simple backchannels. The resulting cascade of false positives severely disrupts conversational flow, culminating in the highest APT of 3.627s. This outcome confirms that relying solely on acoustic energy is an inadequate foundation for intelligent interruption handling.

\textbf{End-to-End Model (Moshi): Robust but Prohibitively Slow.} At the opposite end of the spectrum, the Moshi model embodies an ``overly cautious" strategy. It demonstrates remarkable robustness against backchannels, achieving the lowest average FIR 0.118 and a near-perfect 0.010 on the English set. However, this accuracy is achieved at the expense of cripplingly slow responsiveness. With an average IRL of 2.517s, it is the slowest model, making its latency unacceptable for fluid conversation. This sluggishness translates to a high APT of 3.192s, rendering the model effective in theory but impractical in application.

\textbf{Optimized VAD Models: Incremental but Insufficient Improvements.} The models featuring enhanced VADs offer incremental gains but ultimately fail to resolve the core challenge. While Freeze-Omni strikes a good trade-off than  to achieve the lowest APT among the baselines (2.129s), its FIR remains prohibitively high at 0.532. Similarly, FireRedChat's pVAD shows promise in controlled conditions, exhibiting strong robustness to non-speech sounds (FIR of 0.098 on the Noise/Silence set). However, it falters in realistic dialogue, where its FIR is still high and its IRL climbs to 1.045s. These results indicate that such optimizations, while beneficial for noise suppression, do not imbue the models with the semantic understanding required to differentiate interruption intent from simple backchannels in complex conversations.

\textbf{Our Model: Achieving the Optimal Balance.} In stark contrast, our proposed model decisively outperforms all baselines by successfully resolving the core responsiveness-robustness trade-off. It achieves the lowest average APT of 0.711s—a nearly threefold reduction compared to the best-performing baseline. This superior performance is not limited to conversational data; it extends to challenging acoustic conditions. In the Noise/Silence tests, our model demonstrates exceptional resilience, posting a near-zero FIR of 0.026. This result highlights not only the model's accuracy but also its robustness and generalizability against non-semantic acoustic interference, a critical capability where other systems falter.

\subsection{Qualitative Analysis}
While the quantitative results in Table \ref{tab:metrics_comparison} clearly establish our model's superiority, they also reveal two seemingly counter-intuitive outcomes that warrant deeper investigation. This analysis aims to dissect these puzzles, providing a more nuanced understanding of the behaviors of different interruption handling strategies.

\begin{table}[htbp]
\centering
\caption{Comparison of false alarm number and average penalty time for different models under Silence and Noise conditions.}
\label{tab:silence_noise_comparison}
\begin{tabular}{@{}l cc cc@{}}
\toprule
& \multicolumn{2}{c}{\textbf{Silence}} & \multicolumn{2}{c}{\textbf{Noise}} \\
\cmidrule(lr){2-3} \cmidrule(lr){4-5}
\textbf{Model} & FAN \ & APT & FAN \ & APT \\
\midrule
FSMN-VAD & 159 & 5.420 & 37 & 3.846 \\
Freeze-Omni & 67 & 2.286 & 10 & 0.716 \\
FireRedChat & 0 & 0 & 49 & 6.145 \\
Moshi & 168 & 5.720 & 64 & 7.810 \\
SID-model (proposed) & 10 & 0.340 & 3 & 0.028 \\
\bottomrule
\end{tabular}
\end{table}


A key observation is that Moshi, which achieved outstanding robustness on the English conversational dataset, with FIR of 0.010, paradoxically yielded a high overall APT, 3.192s. The root of this discrepancy lies in its performance on the Noise/Silence set. While highly robust to human backchannels, Moshi proved surprisingly sensitive to subtle, non-semantic noises, leading to a high FIR of 0.464 and a catastrophic APT of 7.810s in this condition. This suggests that Moshi's robustness is not rooted in a true semantic understanding of interruption intent. Rather, its mechanism appears to be based on a pattern-matching model that distinguishes human speech from other acoustic patterns. It fails when presented with non-speech sounds that it misinterprets as turn-taking cues, revealing a critical flaw in its approach.


Another puzzling result is that on the Noise/Silence set, FireRedChat incurred a significantly higher APT, 2.458s, than Freeze-Omni, 1.658s, despite achieving a lower overall FIR, 0.098 vs. 0.154. The data in Table \ref{tab:silence_noise_comparison} reveals that FireRedChat's pVAD is highly effective at filtering out the short, low-energy sounds in the Silence clips, triggering FIR fewer alarms than Freeze-Omni. However, it is frequently and incorrectly triggered by the sustained sounds in the long Noise clips. 
Since the penalty for a False Positive is defined by the total remaining duration of the current turn, a single error in a long-duration noise clip incurs a much heavier penalty than multiple errors in short silence segments. Therefore, although FireRedChat commits fewer total false alarms, its failure to handle sustained noise disproportionately inflates its APT. This observation highlights the APT metric's ability to penalize ``catastrophic" errors—those that would lead to prolonged system malfunction—more heavily than transient glitches, providing a more ecologically valid measure of system reliability.

\section{Conclusion}
In this paper, we address the critical challenge of intelligent interruption handling in spoken dialogue systems by introducing SID-Bench, a novel real-world benchmark, and a corresponding evaluation framework centered on the APT metric. We also propose a novel LLM-based interruption detection model designed to understand user intent at a semantic level. Our extensive experiments demonstrate that this model resolves the long-standing trade-off between responsiveness and robustness, achieving a nearly three-fold improvement in APT over the best-performing baseline. This work not only validates our semantic-driven approach but also provides the community with a robust tool for future research. Future work will focus on integrating this module into a comprehensive, fully generative SDS, expanding SID-Bench to support additional languages, and correlating our automated metrics with human perceptual judgments.

\bibliographystyle{IEEEbib}
\bibliography{icme2026}

\end{document}